\documentclass[a4paper]{jpconf}
\usepackage{graphicx}
\usepackage{url}

\newcommand{\enex}{\ensuremath{E^x}}
\newcommand{\popfact}{\ensuremath{\mathcal{P}}}
\newcommand{\crosssec}{\ensuremath{\sigma}}
\newcommand{\sigmav}{\ensuremath{\left<\crosssec v\right>^*}}

\newcommand{\photoratenostar}{\ensuremath{\lambda}}
\newcommand{\photorate}{\ensuremath{\photoratenostar^*}}
\newcommand{\xfact}{\ensuremath{X}}
\newcommand{\massred}{\ensuremath{m_\mathrm{red}}}
\newcommand{\stellcs}{\ensuremath{\crosssec^*}}

\newcommand{\rmA}{\ensuremath{\mathrm{A}}}

\newcommand{\kb}{\ensuremath{k_\mathrm{B}}}

\begin{document}
\title{Monte Carlo variations as a tool to assess nuclear physics uncertainties in nucleosynthesis studies}

\author{Thomas Rauscher}

\address{Department of Physics, University of Basel, Klingelbergstr.\ 82, 4056 Basel, Switzerland}

\ead{Thomas.Rauscher@unibas.ch}

\begin{abstract}
The propagation of
uncertainties in reaction cross sections and rates of neutron-, proton-, and $\alpha$-induced reactions into the final 
isotopic abundances obtained in nucleosynthesis models is an important issue in studies of nucleosynthesis and Galactic 
Chemical Evolution. 
We developed a Monte Carlo method to allow large-scale postprocessing studies of the impact of nuclear 
uncertainties on nucleosynthesis.
Temperature-dependent rate uncertainties combining realistic experimental and theoretical 
uncertainties are used. From detailed statistical analyses uncertainties in the final abundances are derived as
probability density distributions. Furthermore, based on rate and abundance correlations an 
automated procedure identifies the most important reactions in complex flow patterns from superposition of many zones or tracers.
The method already has been applied to a number of nucleosynthesis processes.
\end{abstract}

\section{Introduction}
Low-energy reaction cross sections with light projectiles are required to determine astrophysical reaction rates and to constrain 
the production of 
nuclides in various astrophysical environments. Off stability only theoretically predicted 
reaction rates are used in nucleosynthesis calculations, both for neutron-rich and proton-rich nuclides.
Even along stability not all 
rates can be constrained experimentally and 
combinations of experimental data and nuclear theory have to be used. Our studies 
address an important question in the context of astrophysical applications: how 
uncertainties in cross sections of reactions induced by neutrons, protons, and $\alpha$-particles 
propagate into the final 
isotopic abundances obtained in nucleosynthesis models. 
This information is important for astronomers to interpret their observational data, for groups studying the enrichment of the 
Galaxy 
over time with heavy elements, and in general for disentangling uncertainties in nuclear physics from those in the astrophysical 
modelling. To this end, over several years we developed a Monte Carlo (MC) method to allow large-scale studies of the impact of 
nuclear uncertainties on nucleosynthesis.
The MC framework \textsc{PizBuin} uses an efficient, parallelised reaction network solver, allowing to postprocess 
astrophysical trajectories with a large reaction network including several thousand nuclides and several tens of thousand 
reactions. The trajectories specify the temporal 
evolution of density and temperature and can be taken from any astrophysical simulation. Further advantages of the 
\textsc{PizBuin} code are: (i) the analysis can be 
performed combining many such trajectories, for example, for trajectories describing different regions of an exploding star; (b) 
temperature-dependent uncertainties are used, reflecting the special conditions inside a stellar plasma; (c) correlations between 
rate and final abundance variations are exploited to automatically identify key reactions.

\section{Temperature-dependent uncertainties in thermonuclear reaction rates}

To understand the philosophy underlying the temperature-dependent variation factors used in the MC procedure it is 
necessary to recall some details of the astrophysical reaction rates. A reaction rate for an interaction between two reaction 
partners a and A in a plasma with temperature $T$ is defined as
\begin{equation}
\label{eq:stellrate}
r^*(T) = \frac{n_\mathrm{a} n_\mathrm{A}}{1+\delta_\mathrm{aA}} R^*(T) = \frac{n_\mathrm{a} n_\mathrm{A}}{1+\delta_\mathrm{aA}} 
\sigmav (T) = \frac{n_\mathrm{a} n_\mathrm{A}}{1+\delta_\mathrm{aA}} \int_0^\infty \stellcs(v,T) v \Phi(v,T) \,\rmd v\quad,
\end{equation}
where $\Phi$ is the appropriate distribution of the relative velocities in the stellar plasma. For nuclei and stellar temperatures 
this is a Maxwell-Boltzmann distribution. The quantity $R^*=\sigmav$ is called stellar reactivity and $n_\mathrm{a}$, 
$n_\mathrm{A}$ are the number densities of projectile and target, respectively. Alternatively, Eq.\ (\ref{eq:stellrate}) could be 
written as integration over c.m. energy, see below.

An asterisk indicates quantities adapted for stellar plasma effects. An 
important effect for intermediate and heavy nuclei is the thermal excitation, i.e., a fraction of the nuclei will be in excited 
states $\mu$ according to a population coefficient $\popfact_\mu(T)$. Thus, the stellar reactivity includes a sum over reactions 
proceeding on excited states (starting from the ground state with $\mu=0$),
\begin{eqnarray}
R^*=\sigmav_\mathrm{aA}&=& \popfact_0 R_0 + \popfact_1 R_1 + \popfact_2 
R_2 +
\dots
= \sum_\mu \popfact_\mu R_\mu =\nonumber \\
&=&\left[\frac{8}{\massred^\mathrm{aA} \pi}\right]^\frac{1}{2}
[\kb T]^{-\frac{3}{2}} \sum_\mu {\left[ \popfact_\mu \int_0^\infty {\crosssec^\mu_\mathrm{aA} E_\mu^\rmA
\rme^{-\frac{E_\mu^\rmA}{\kb T}}\,\rmd E_\mu^\rmA}\right] }\quad.\label{eq:stellrate1}
\end{eqnarray}
This means that projectiles with Maxwell-Boltzmann distributed energies are acting on each level $\mu$ separately. Since each
excited state is exposed
to the full energy range, each integral in Eq.\ (\ref{eq:stellrate1}) has its own energy scale, ranging from zero to infinity but
shifted
relative to each other by the excitation energy of the level $\enex_\mu$ (note the different $\rmd E_\mu^\rmA$ in each integral of
the sum). This can be seen more easily when explicitly inserting the population coefficient, leading to \cite{fowler,intj}
\begin{eqnarray}
\sum_\mu {\left[ \popfact_\mu \int_0^\infty {\crosssec^\mu_\mathrm{aA} E_\mu^\rmA \rme^{-\frac{E_\mu^\rmA}{\kb T}}\,\rmd
E_\mu^\rmA}\right] }
&=& \sum_\mu \int_0^\infty { \popfact_\mu \crosssec^\mu_\mathrm{aA} E_\mu^\rmA \rme^{-\frac{E_\mu^\rmA}{\kb T}} \,\rmd E_\mu^\rmA} =
\nonumber \\
 = \sum_\mu \int_0^\infty { \frac{g_\mu^\rmA \rme^{-\frac{\enex_\mu}{\kb T}}}{g_0^\rmA G^\rmA_0} \crosssec^\mu_\mathrm{aA}
E_\mu^\rmA
\rme^{-\frac{E_\mu^\rmA}{\kb T}} \,\rmd E_\mu^\rmA} &=& \frac{1}{G^\rmA_0} \sum_\mu \int_0^\infty {
\frac{g_\mu^\rmA}{g_0^\rmA} \crosssec^\mu_\mathrm{aA}
E_\mu^\rmA \rme^{-\frac{E_\mu^\rmA+\enex_\mu}{\kb T}}}\,\rmd E_\mu^\rmA
, \label{eq:stellrate2}
\end{eqnarray}
where $G^\rmA_0=G^\rmA(T)/g_0$ is the nuclear partition function $G^\rmA(T)$ normalized to the ground state (g.s.) spin factor 
$g_0=2J_0+1$. (The other spin factors above are defined similarly.) Because of this energy shift of
the Maxwell-Boltzmann distributions it is not straightforward to define the quantity $\stellcs (E,T)$ to be used in
Eq.\ (\ref{eq:stellrate}). Mathematically transforming Eq.\ (\ref{eq:stellrate2}) by exchanging summing and integration, and 
providing the appropriate transformation of the integration variable, it is possible \cite{fowler,gscontrib} to derive the quantity 
$\stellcs$ in 
the single integral of Eq.\ (\ref{eq:stellrate}),
\begin{equation}
\sigma^*_\mathrm{aA} (E,T) = \frac{1}{G^\mathrm{A}_0(T)}\sum_\mu \sum_\nu \frac{g_\mu^\rmA}{g_0^\rmA} \frac{E-E^x_\mu}{E}
\sigma^{\mu\nu}_\mathrm{aA} (E-E^x_\mu)\quad.\label{eq:stellcsweight}
\end{equation}
The individual cross sections $\sigma^{\mu\nu}_\mathrm{A}$ for transitions $\mu \rightarrow \nu$ between initial state $\mu$ and 
final state $\nu$ are evaluated at an 
energy
$E-\enex_\mu$, with  $\sigma^{\mu\nu}_\mathrm{A}=0$ for  $(E-E^x_\mu)\leq 0$. Thus, the
weight of an individual excited-state cross section in the stellar cross section is
\begin{equation}
\label{eq:stellarweight}
\mathcal{W}_\mu (E)= \frac{E-E^x_\mu}{E}=1-\frac{E^x_\mu}{E}\quad.
\end{equation}
This is also weighting its contribution to the reaction rate integral. The weight $\mathcal{W}_\mu$ is energy dependent but since 
the
largest contribution of the cross section to the integral comes from around the Gamow energy $E_\mathrm{G}$, the effective weight 
can
be approximated by $\mathcal{W}_\mu(E_\mathrm{G})$. Similar considerations apply to 
photodisintegration reactions
but a slightly different weight
is obtained
because shifted Planck distributions do not fully cancel with the population factors \cite{book}.

According to Eq.\ (\ref{eq:stellrate1}),
the contribution $X_\mu$ of a level $\mu$ to the stellar rate or reactivity can be calculated using
\begin{equation}
\label{eq:stellcontrib}
\xfact_\mu = \frac{\popfact_\mu R_\mu}{R^*}=\frac{g_\mu \rme^{-\frac{\enex_\mu}{\kb
T}}}{G} \frac{R_\mu}{R^*}
= \frac{g_\mu}{g_0} \frac{1}{G_0} \rme^{-\frac{\enex_\mu}{\kb T}} \frac{R_\mu}{R^*} \quad.
\end{equation}
It also follows from Eq.\ (\ref{eq:stellrate1}) that
\begin{equation}
\sum_\mu {\xfact_\mu} = 1 \; \mathrm{with} \; 0\leq \xfact_\mu \leq 1\quad.
\end{equation}
The same applies for photodisintegration reactions with their stellar rates $\photorate$,
\begin{equation}
\label{eq:stellcontribphoto}
\xfact_\mu = \frac{\popfact_\mu \photoratenostar_\mu}{\photorate}=\frac{g_\mu \rme^{-\frac{\enex_\mu}{\kb
T}}}{G} \frac{\photoratenostar_\mu}{\photorate}
= \frac{g_\mu}{g_0} \frac{1}{G_0} \rme^{-\frac{\enex_\mu}{\kb T}} \frac{\photoratenostar_\mu}{\photorate} \quad.
\end{equation}
For the ground-state contribution ($\mu=0$) this reduces to
\begin{eqnarray}
\xfact_0 (T) &=& \frac{1}{G_0 (T)} \frac{R_0}{ R^* (T)} \quad, \label{eq:gscontrib}\\
\xfact_0 (T) &=& \frac{1}{G_0 (T)} \frac{\photoratenostar_0}{ \photorate (T)} \quad.
\label{eq:gscontribphoto}
\end{eqnarray}
Obviously, the combined contribution $\xfact_\mathrm{exc}$ of reactions on all excited states (not including the ground state) is 
given by
\begin{equation}
\xfact_\mathrm{exc} (T) = 1 - \xfact_0 (T)\quad.
\end{equation}

\begin{figure}[tbh]
\begin{center}
\includegraphics[width=0.53\columnwidth]{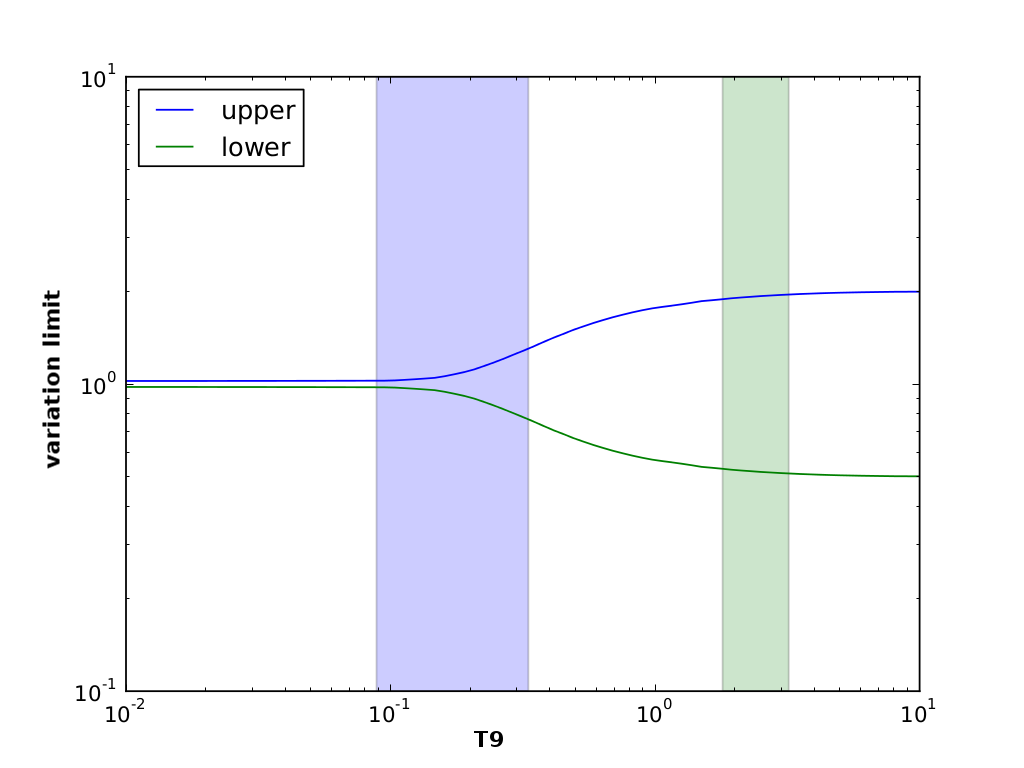}
\includegraphics[width=0.46\columnwidth]{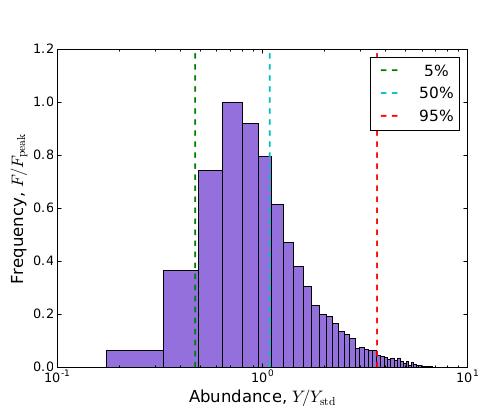}
\end{center}
\caption{Left panel: Example of the temperature-dependent variation range. Shown are the upper and lower limits of the variation 
factor of the rate for $^{157}$Gd(n,$\gamma$)$^{158}$Gd as function of plasma temperature $T_9$ in 10$^9$ K. Shaded areas indicate 
the relevant $T_9$ ranges for the s-process (left) and the $\gamma$-process (right).
Right panel: Example of the abundance distributions obtained in the MC runs. The bounds encompassing 5\% and 95\% of the 
distribution are marked to allow to use them as uncertainty measures.\label{fig:varfact}}
\end{figure}

Due to elevated temperatures in late stellar burning phases and in explosive nucleosynthesis most nuclei, except for the lightest 
ones, have their excited states populated. When using measured cross sections and their uncertainties, it has to be considered that 
an experiment may not have constrained reactions on all populated excited states. Most of the reaction data available only concerns 
target nuclei being in the ground state. The g.s. contribution in explosive burning of intermediate to heavy nuclei with high level 
density can be a few percent only (a few permille for photodisintegrations) \cite{gscontrib,advances}. Therefore the actual 
uncertainty in the stellar 
rate must contain a temperature-dependent combination of experimental and theory uncertainties, just as the stellar rate is given 
by a superposition of reactions on various target states. Using the g.s.\ contribution $X_0(T)$ to the stellar 
rate, the total uncertainty factor 
$u^*(T)$ of a reaction rate can be given as \cite{advances,apjlett}
\begin{equation}
u^*(T)=U_{\mathrm{exp}}+\left[ U_\mathrm{th}-U_{\mathrm{exp}} \right] \left[ 1-\xfact_0(T)\right] \label{eq:uncertainty} \quad,
\end{equation}
where $U_{\mathrm{exp}}$ and $U_\mathrm{th}$ are the experimental and theoretical uncertainty factors, respectively, and
$U_\mathrm{th}>U_{\mathrm{exp}}$. Assuming a symmetric uncertainty, this would limit the range of rate variation factors to $u^*$ 
and $1/u^*$. An example for the $T$ dependence of these limits is shown in Fig.\ \ref{fig:varfact} (left) for the reaction rate of 
$^{157}$Gd(n,$\gamma$)$^{158}$Gd. Although the ground-state cross section is tightly constrained experimentally \cite{bao}, 
reactions on excited states of $^{157}$Gd contribute significantly to the stellar rate at increased temperatures. In this case, 
theoretical uncertainties start to become important already at s-process temperatures and they dominate at typical $\gamma$-process 
temperatures.

\section{Monte Carlo variations}

In our MC calculations different uncertainty limits were assigned to different reaction types, with the temperature dependence 
obtained from Eq.\ (\ref{eq:uncertainty}). Experimental 2$\sigma$ uncertainties $U_{\mathrm{exp}}$ were considered for g.s.\ 
contributions 
when available. Theoretical uncertainties $U_\mathrm{th}$ for g.s.\ and excited state 
contributions were assigned symmetric or asymmetric uncertainties, as appropriate, which are assumed to include systematic errors. 
In particular, predicted rates for neutron-induced reactions received an uncertainty limit of a factor of 2 (0.5), whereas an 
asymmetric uncertainty was used for predicted rates involving protons (factor of 2.0 for upper and 0.33 for lower limit) and 
$\alpha$ particles (factor 2.0 up and 0.1 for the lower limit). The same variation factor is used for forward and reverse rate 
because \textit{stellar} rates obtained with thermally 
populated excited states are connected by detailed balance. Therefore, changing the reactivity in one reaction direction affects 
the reactivity in the reverse direction in the same manner.

The MC variation factors provided by the random number generator are values $0\leq f_\mathrm{MC}\leq 1$, drawn from a uniform 
distribution. The actual varied rate $r^*_\mathrm{MC}(T)$ is computed from
\begin{equation}
r^*_\mathrm{MC}(T)=r_{\mathrm{lo}}(T)+f_\mathrm{MC}\left[ r_{\mathrm{hi}}(T)-r_{\mathrm{lo}}(T)\right] \quad.\label{eq:rate}
\end{equation}
The upper and lower rate limits $r_{\mathrm{hi}}=u^*(T)r^*_\mathrm{std}(T)$ and $r_{\mathrm{lo}}=r^*_\mathrm{std}(T)/u^*$ are 
derived from the standard rates $r^*_\mathrm{std}$ and the uncertainty limits from Eq.\ (\ref{eq:uncertainty}). Note 
that they depend on $T$ and can be asymmetric. It is also important to note that $f_\mathrm{MC}$ does not depend on $T$ as otherwise 
this would result in non-analytic rates. We have tested how many variations are needed to obtain a statistically meaningful sample 
of a rate within its uncertainty and found that several 1000 variations are sufficient. Our studies use 10000 MC iterations, each 
using a separate random variation factor for a rate. Therefore each rate is varied 10000 times.
In this approach the required computational time is largely independent of the number of varied reaction rates. 
Rather, it is determined by the time taken to follow the reaction network through a given trajectory and the number of considered 
trajectories. The solution time for the network is determined by the number of nuclides included and 
the number of reactions connecting them.

Figure \ref{fig:varfact} shows an example of the binned uncertainty distribution in a final abundance after 10000 
iterations. It results from the 
combined uncertainties of all contributing rates. The more reactions are contributing, the closer the distribution shape will be to 
a lognormal distribution, regardless of the distribution type (uniform, Gaussian, \dots) of the individual rate uncertainties.

\section{Determination of key reactions}

The individual variation of a single reaction only tests the sensitivity of a particular final abundance to changes in this rate. 
It does not provide information on the actual importance of this rate in the final abundance because uncertainties in many rates 
may contribute. The advantage of the MC approach is that all rates are varied simultaneously and therefore the combined impact of 
all rates on an abundance is obtained. Using the MC data to extract correlations between the variation of a given rate with the 
variation of an abundance \textit{while having varied all other rates as well} allows to determine the actual key reactions, i.e., 
those reactions which contribute most to the uncertainty in the final abundance. A better constraint of these reactions will lead 
to a reduction of the abundance uncertainty. Methods to quantify correlations can be categorized 
into rank methods and product-moment methods \cite{kendall55,mathguru}. Rank correlation methods, although 
formally assumed to better account for data outliers, are losing
information in the ranking procedure and are rather unsuited for the purpose of correlating reactions and abundances. For our 
particular application they also do not allow to make easily a weighted combination of the contribution of several trajectories to 
a total uncertainty.
We found the Pearson product-moment correlation coefficient to be more suitable
to quantify correlations \cite{pearson}. Data outliers to which the Pearson coefficient would be 
vulnerable do not appear in an
analytic variation of reaction rates.

Since we are interested in key rates which globally affect the final abundances and not just those in one trajectory, it was 
necessary to modify the basic Pearson formula to provide a weighted average over all trajectories used. Our weighted 
correlation coefficient $r^q_\mathrm{corr}$ is given by \cite{tommy}
\begin{equation}
\label{eq:weightcorr}
r^q_\mathrm{corr}=\frac{\sum_{ij} {^qw_{j}^2} \left( f_{ij} -\overline{f}_j \right)\left( ^qY_{ij} - \overline{^qY}_j\right) 
}{\sqrt{\sum_{ij} {^qw_j^2}\left( f_{ij} -\overline{f}_j \right)^2}\sqrt{\sum_{ij} {^qw_j^2} \left( ^qY_{ij} -\overline{^qY}_j 
\right)^2}} \quad.
\end{equation}
The trajectory is identified by $j$ and the iteration by $i$, with variation factor $f_{ij}=r^*_\mathrm{MC}/r^*_\mathrm{std}$ of 
the rate and final abundance $^qY_{ij}$ of nuclide $q$ resulting from this variation. The barred quantities are the means of the 
samples of variation factors $\overline{f}_j=\left(\sum_{i=1}^k f_{ij} \right)/k$ and abundances 
$\overline{^qY}_j=\left(\sum_{i=1}^k {^qY_{ij}} \right)/k$ with respect to the number of MC iterations $k$.
To connect all rates to all abundances of 
interest, the number of weighted correlation factors to be computed for each nuclide of interest is the number of reactions in the 
network. The weight of each trajectory is calculated from the relative abundance change
\begin{equation}
\label{eq:normprod}
^qw_j=\frac{|^qY^\mathrm{std}_j-^qY^\mathrm{ini}_j|}{\sum_j {|^qY^\mathrm{std}_j-^qY^\mathrm{ini}_j}|}
\end{equation}
for each nuclide $q$ with initial abundance $^qY^\mathrm{ini}_j$ in trajectory $j$. The final abundances obtained with the standard 
rate set are denoted by $^qY^\mathrm{std}_j$.

Positive values of the Pearson coefficients $-1\leq r^q_\mathrm{corr}\leq 1$ 
indicate a direct correlation between rate change and abundance change, whereas negative values signify an inverse correlation, 
i.e., the abundance decreases when the rate is increased. Larger absolute values $|r^q_\mathrm{corr}|$ indicate a stronger 
correlation and this can 
be used for extracting the most important reactions from the MC data. We define a key rate (i.e., a rate dominating the final 
uncertainty in the production of a nuclide)
by having $|r^q_\mathrm{corr}|\geq 0.65$.

\section{Conclusion}

The method so far was already applied to a number of processes: the 
$\gamma$ process ($p$ process) \cite{tommy} and the $\nu p$ process \cite{mcnup} in core-collapse supernovae (ccSN), 
the production of $p$ nuclei in white dwarfs exploding as 
thermonuclear (type Ia) supernovae \cite{snIa} (SNIa), the weak $s$ process in massive stars \cite{nobuya}, and the main $s$ 
process in 
AGB stars \cite{gabriele}. Included in the MC variations were only reactions on isotopes of Fe and heavier elements. These meet the 
chosen assumptions on nuclear uncertainties of non-resonant compound reactions. A few selected reactions on lighter nuclides of 
particular interest were individually varied.

As a general observation, it was found that the production uncertainties for 
most nuclei are below factors of $2-3$, despite of theoretically predicted rates dominating nucleosynthesis at high temperature 
with much larger assumed uncertainties. This indicates that larger uncertainties of individual rates can cancel out. Another way to 
reduce the dependence on individual rate uncertainties is an adaptation of the reaction flow depending on the values of individual 
rates. This is accompanied by a lack of key reactions because bottlenecks in the flow are avoided. Both effects would not be seen 
or underestimated in a manual variation of only a single or few rates. Both effects lead to comparatively small uncertainties, 
especially when compared to the uncertainties inherent in the astrophysical modelling of the nucleosynthesis processes.

Except 
for the $s$ process, 
particular conditions in nucleosynthesis trajectories and sometimes even the contributions of nucleosynthesis sites (e.g., for the 
$\gamma$-process) often are not well constrained by astrophysical simulations and lead to much larger 
(systematic) uncertainties in 
the production level of a particular nuclide (or whether it is produced at all). For details, see the cited literature 
\cite{tommy,mcnup,snIa,nobuya,gabriele}.

In summary, we have provided a powerful, flexible method to assess nucleosynthesis uncertainties originating from uncertainties in 
the astrophysical reaction rates. In future investigations, we plan to apply the method to further nucleosynthesis processes and 
also to study the impact of correlations between rates.

\section{Acknowledgments}
I thank the many contributors involved in developing and testing the MC and analysis codes, especially N. Nishimura, G. Cescutti, 
U. Frischknecht. C. Winteler, J. Reichert, R. Hirschi, A. St.J. Murphy. The work was partially supported by the UK 
STFC (grants ST/M000958/1, ST/M001067/1), the EU ERC (GA 321263-FISH; EU-FP7-ERC-2012-St Grant 306901), the EU COST action 
CA16117 (ChETEC), the Royal Society, and the Swiss NSF.


\section*{References}
\begin{thebibliography}{9}
\bibitem{fowler} Fowler W A 1974 \textit{Quart.\ J. Roy.\ Astron.\ Soc.} {\bf 15} 82
\bibitem{intj} Rauscher T 2011 \textit{Int.\ J. Mod.\ Phys.\ E} {\bf 20} 1071
\bibitem{gscontrib} Rauscher T 2012 \textit{Ap.\ J. Suppl.} {\bf 201} 26
\bibitem{book} Rauscher T 2020 \textit{Essentials for Nucleosynthesis and Theoretical Nuclear Astrophysics} (Bristol: IOP 
Publishing)
\bibitem{advances} Rauscher T 2014 \textit{AIP Advances} {\bf 4} 041012
\bibitem{apjlett} Rauscher T 2012 \textit{Ap.\ J. Lett.} {\bf 755} L10; \textit{Ap.\ J. Lett.} {\bf 864} L40
\bibitem{bao} Bao Z Y, et al 2000 \textit{ADNDT} {\bf 76} 70
\bibitem{kendall55} Kendall M G 1955 \textit{Rank correlation methods} (London: Charles Griffin)
\bibitem{mathguru} Hemmerich W A 2017 \textit{Korrelation, Korrelationskoeffizient}\\
online at 
\url{https://web.archive.org/web/20170624012731/http://matheguru.com/stochastik/korrelation.html}
\bibitem{pearson} Pearson K 1895 \textit{Proc.\ Roy.\ Soc.\ Lon.} {\bf 58} 240
\bibitem{tommy} Rauscher T, Nishimura N, Hirschi R, Cescutti G, Murphy A StJ and Heger A 2016 \textit{MNRAS} {\bf 463} 4153
\bibitem{mcnup} Nishimura N, Rauscher T, Hirschi R, Cescutti G, Murphy A StJ and Fr\"ohlich C 2019 \textit{MNRAS} {\bf 489} 1379
\bibitem{snIa} Nishimura N, Rauscher T, Hirschi R, Murphy A StJ, Cescutti G and Travaglio C 2018 \textit{MNRAS} {\bf 474} 3133
\bibitem{nobuya} Nishimura N, Hirschi R, Rauscher T, Cescutti G and Murphy A StJ 2017 \textit{MNRAS} {\bf 469} 1752
%\bibitem{gabriele} Cescutti G, Hirschi R., Nishimura N, den Hartogh J W, Rauscher T, Murphy A StJ and Cristallo S 2018 
\bibitem{gabriele} Cescutti G, et al 2018
\textit{MNRAS} {\bf 478} 4101
\end{thebibliography}
\end{document}